\documentclass[12pt,aps,floats,showpacs,amssymb,tightenlines]{revtex4-2}

\usepackage{color}
\usepackage{graphicx}
\usepackage{amsmath}
\usepackage{amsfonts}
\usepackage{amscd}
\usepackage{epsfig}
\usepackage{amssymb}
\usepackage{tabularx}

\begin{document}

\title{An experiment to test the isotropy of the one-way speed of light.}
\author{Reinaldo J. Gleiser} \email{gleiser@famaf.unc.edu.ar}

\affiliation{Instituto de F\'{\i}sica Enrique Gaviola and FAMAF,
Universidad Nacional de C\'ordoba, Ciudad Universitaria, (5000)
C\'ordoba, Argentina}

\begin{abstract}
An experimental setup capable, in principle, to test the isotropy of the {\em one 
way} propagation of light to 1 part in $10^8$ (or better), is suggested.
\end{abstract}

\pacs{04.20.Jb}

\maketitle

\section{Introduction.}

The speed of light in a vacuum is one of the most fundamental quantities in 
modern physics. It is well known, however, and there has been ample debate in 
the literature, that while the {\em two way} speed of light is experimentally 
well defined, the {\em one way} speed of light cannot be measured as long as 
speed is conceived as the ratio of distance travelled to time taken for this 
travel \cite{refes}. We shall call this concept of speed as ``ballistic''. The 
problem stems mainly from the fact that the ``ballistic'' measurement requires the 
presence of synchronized clocks at each end of the path travelled, and this 
synchronization uses in turn light signals, creating a circular argument that 
so far has made this type of measurement invalid.

Going back to the measurement of the {\em two way} speed of light, in the 
typical experiment light is sent from some initial point to a distant mirror, 
where it is reflected so that it returns to the initial point. By measuring the 
distance $L$ to the mirror, and the time $T$ taken for the full trip (which 
requires only {\em one} clock, i.e. no synchronization), the speed is defined as,
\begin{equation}
\label{velo1}
c=\frac{2 L}{T}
\end{equation}
 
In principle, $c$ in (\ref{velo1}) represents only the {\em average} speed of 
light in the two way trip. This definition contains the implicit assumption that 
the speed of light is independent of the direction of propagation, that is, 
that the {\em one way} speed of light is {\em isotropic}, but this is precisely 
what the ``ballistic'' method cannot establish.  

We remark that the ``ballistic'' idea of measuring the speed of light has no 
consideration of the fact that the propagation of light is fundamentally a {\em 
wave} phenomenon. As such, in a vacuum, in the absence of dispersion, the speed 
of propagation of a monochromatic wave of period $\tau$, and wavelength 
$\lambda$ is,
\begin{equation}
\label{velo2}
c=\frac{\lambda}{\tau}
\end{equation}
Therefore, we may compute $c$ if we can measure both $\tau$, and $\lambda$. 
This seems to take us back to the ballistic idea, but we shall argue that this 
is not the case. Assume we have a plane monochromatic wave propagating in a 
given direction. By placing a (single) clock at some fixed point, anywhere in 
the path of the plane wave, we can measure $\tau$. Assume further that after 
the measurement of $\tau$ the wave goes through some fixed structure, and that 
this gives rise to a certain diffraction pattern, that can be observed further 
along the path of the wave. The fundamental point here is that the diffraction 
pattern is completely determined by the geometry of the structure and the 
wavelength of the diffracted light, with {\em no} measurement of time involved, 
or distance travelled. Thus, if the geometry of the structure is known, a 
measurement of the diffraction pattern provides immediately a measure of the 
wavelength $\lambda$. Therefore, using the {\em wave} properties of the 
propagation of light we can obtain $c$ in (\ref{velo2}), with {\em no} 
synchronization involved.  

In practice, $c$ in (\ref{velo1}) can be measured to much larger accuracy than 
what (\ref{velo2}) could provide. However, as just discussed, the two values of 
$c$ should coincide if the isotropy assumption is valid. As it turns out, and 
will be the main subject of this paper, the ideas behind (\ref{velo2}) can be 
better used, not to obtain $c$, but rather to test the {\em isotropy} of $c$. 
As will be discussed in what follows, this does not even require measuring 
times or distances, as only intensities will be involved. As we shall show, 
even a simple setup could be used to test this isotropy to the order of 1 part 
in $10^7$ or $10^8$, and possibly, even better.

In the next Section we review the basics of the diffraction of light by a 
diffraction grating. Then, in Section III we present our basic idea, which, as 
explained there, is based on the interference of two separate monochromatic but 
coherent diffracted beams, which are setup in such a way that they move in 
opposite directions as the wavelength of the diffracted light changes. In 
Section IV we discuss several concrete examples, assigning definite numerical 
values to the different parameters involved, and show that the setup could, in 
principle, display a sensitivity to changes in $\lambda$, and therefore, for 
fixed $\tau$, to $c$, in the order of one part in $10^8$. We discuss a sketch 
of what could be a concrete experimental arrangement in Section V. We close 
with some final comments in Section VI. 

We have also added an Appendix with two Sections, where we consider a slightly 
more general case regarding the experimental parameters and show, both 
theoretically and with an explicit example, that the essential results 
regarding the sensitivity of the proposed setup are not modified in this case.

The final conclusion is that this type of experimental arrangement might 
provide either an upper bound to a possible anisotropy in the one way 
propagation, and therefore a measure of the one way speed of light to that 
accuracy, or, on the contrary, establish the existence of an anisotropy with a 
precision of a few meters per second. In either case, the results would 
undoubtedly be relevant as regards the physical properties of space time.

\section{The diffraction grating.}

We review briefly the basics of the diffraction of light by a
diffraction grating \cite{BornWolf}. Consider the situation shown in
Figure 1. 
 
\begin{figure}[h!] 
\vspace{0cm}
\centerline{\includegraphics
[height=7cm,angle=0.0]
{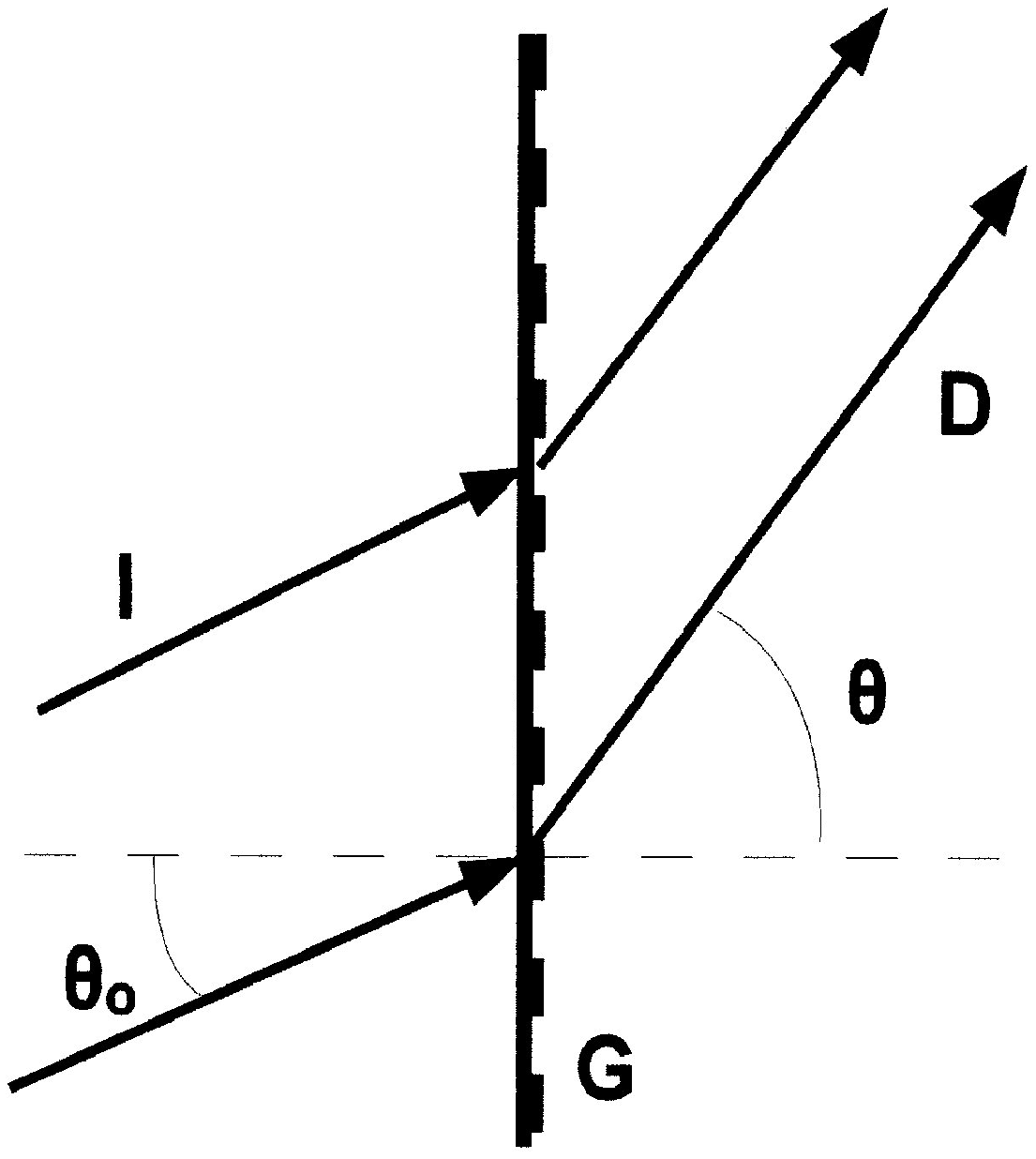}}
\vspace{0cm} \caption{A plane monochromatic wave {\bf I} is incident on 
a diffraction grating {\bf G} at an angle $\theta_0$.
{\bf D} is the diffracted wave emerging at an angle $\theta$.}
\end{figure}

A plane wave of monochromatic light of frequency $\omega= 2\pi/\tau$
is incident at an angle $\theta_0$ on a plane (transmission)
diffraction grating $G$, with a spacing $d$ between grooves. The
complex amplitude $U(p)$ diffracted at a large distance from the
grating, and at an angle $\theta$ from the normal to the plane of
the grating is given by,
\begin{equation}
\label{eq1}
 U(p)=U^{(0)}(p) \frac{1-e^{-iNkdp}}{1-e^{-ikdp}}
\end{equation}
where,
\begin{eqnarray}
\label{eq2}
p&=& \sin\theta -\sin \theta_0, \nonumber \\
k & = & 2 \pi/\lambda,
\end{eqnarray}
with $\lambda$ the wave length corresponding to $\tau$, and we
have omitted an overall factor $e^{i (\omega t +\alpha)}$, with
$\alpha$ a constant. $N$ is the total number of grooves illuminated by the incident
plane wave, $U^{(0)}(p)$ depends on the shape of the grooves, but
does not depend on $N$, and is, in general, slowly dependent on $p$.

Let us assume for simplicity that $\theta_0=0$. Then, the intensity
of the diffracted light in the direction $\theta$, $I(p)=|U(p)|^2$,
can be written as,
\begin{equation}
\label{eq3}
 I(p)=|U^{(0)}(p)|^2
 \left(\frac{\sin(Nkd\sin(\theta)/2)}{\sin(kd\sin(\theta)/2)}\right)^2
\end{equation}

For large $N$, (and fixed $d$ and $\lambda$), this intensity has
sharp maxima with $I(p)=|U^{(0)}(p)|^2 N^2$  for $\theta=\theta_n$,
such that,
\begin{equation}
\label{eq4}
   kd\sin(\theta_n)=2 n \pi
\end{equation}
where $n \neq 0$ is a positive or negative integer, or,
\begin{equation}
\label{eq5}
   \sin(\theta_n)=n\frac{\lambda}{d}
\end{equation}

Similarly, the intensity vanishes for $\theta_m\neq \theta_n$, such
that,
\begin{equation}
\label{eq7}
   \sin(\theta_m)=m\frac{\lambda}{N d}
\end{equation}
where $m \neq 0$ is a positive or negative integer.

Restricting to $n=\pm 1$, and calling $\Delta \theta$ the change in
$\theta$ from the maximum to the closest zero, we have,
\begin{equation}
\label{eq8}
   \sin(\theta+\Delta \theta)=\frac{\lambda}{d} \frac{N+1}{N}
\end{equation}

This can also be written as,
\begin{equation}
\label{eq9}
   \sin(\theta+\Delta \theta)-\sin(\theta)=\frac{\lambda}{Nd}
\end{equation}

Suppose now we have a second normally incident monochromatic plane
wave of frequency $\omega'$, and wavelength $\lambda'$. The
diffracted intensity will then be given by,
\begin{equation}
\label{eq10}
 I'(p)=|{U'}^{(0)}(p)|^2
 \left(\frac{\sin(Nk'd\sin(\theta)/2)}{\sin(k'd\sin(\theta)/2)}\right)^2
\end{equation}
and there will be a maximum of $I'(p)$ for
$\sin(\theta')=\lambda'/d$. If we define $\Delta
\lambda=\lambda'-\lambda$, and $\Delta \theta'=\theta'-\theta$, we
then have,
\begin{equation}
\label{eq11}
   \sin(\theta+\Delta \theta')-\sin(\theta)=\frac{\Delta \lambda}{d}
\end{equation}

If we apply the criterion that the maxima for $\lambda$ and for
$\lambda'$ are seen as separate if the maximum of $\lambda'$
coincides with the closest zero of $\lambda$, we have that $\Delta
\theta'$ must be equal to $\Delta \theta$ of (\ref{eq9}), and
therefore,
\begin{equation}
\label{eq12}
   \frac {\lambda}{\Delta \lambda} = N
\end{equation}
This relation defines the {\em resolving power} of the grating.
Typically, $N$ can be of the order of $10^4$ (or higher), so that
one can resolve wavelengths $\lambda$, and $\lambda'$ that differ on
the order of one part in $10^4$ (or more).

But here we want to consider this set up from a different point of
view. We first notice that for a monochromatic wave we have,
\begin{equation}
\label{eq13}
    \lambda = \frac{2 \pi}{\omega} c
\end{equation}
where $c$ is the speed of propagation of the wave in the direction
considered.

Suppose now the source of the monochromatic plane wave has a given,
fixed frequency $\omega$, and that we point the whole set up in
different directions in space, while keeping the dimensions,
relations and relative orientations of its different components
(source, grating, supports, etc.) fixed. Notice that $c$ in
(\ref{eq13}) corresponds to what is called the {\em one way} speed
of light, as no closed loops are involved. For {\em fixed} $\omega$,
and the rest of the parameters of the set up, a change in $c$ with
the direction in which the set up points, will result in a change in
$\lambda$, and, as a consequence, in the diffraction angle $\theta$.
Therefore, the set up will be sensitive to a possible dependence of
$c$ on the {\em direction} of propagation, and thus provide a test
of the isotropy of the {\em one way} speed of light. How sensitive
can this set up be? In accordance with (\ref{eq12}) and
(\ref{eq13}), and assuming that the same criterion is applicable, we
would have that a {\em negative} result (no observable change in
$c$), would indicate,
\begin{equation}
\label{eq14}
      \left|\frac{\Delta c}{c} \right| < \frac{1}{N}
\end{equation}

But, in deriving and applying (\ref{eq12}) it was implicitly assumed that the 
sources of $\lambda$, and $\lambda'$ were {\em incoherent}. As we shall 
indicate in the next section, where we propose our basic idea for testing the 
isotropy of the one way speed of light, this can be very different when we 
consider a superposition of {\em coherent} plane waves.

\section{The basic idea}

To clarify the basic idea, we refer to Figure 2, where we indicate the $X-Y$ 
plane of an auxiliary orthogonal coordinate system with axes $(X,Y,Z)$. The $Z$ 
axis is perpendicular to the $X-Y$ plane. Two identical plane transmission diffraction 
gratings, $G_1$, and $G_2$ are placed with their grooves parallel to the $Z$ 
axis, with one of their ends touching at the origin of coordinates. The planes 
of the gratings are set at angles $\alpha$, and $-\alpha$ respectively to $Y-Z$ 
plane. Two monochromatic, coherent, plane waves, $I_1$, and $I_2$, are normally 
incident respectively on $G_1$, and $G_2$, giving rise to diffracted amplitudes 
$U_1$, and $U_2$. The total diffracted amplitude, resulting from the 
superposition of $U_1$ and $U_2$ is to be observed at a large distance in the 
direction indicated by the angle  $\phi$. 

Now, in accordance with (\ref{eq1}), for the diffracted amplitudes, in general, we have,
\begin{equation}
\label{eq15}
 U_1(p_1)=e^{i\varphi}
 \frac{1-e^{-iNkd\sin(\theta_1)}}{1-e^{-ikd\sin(\theta_1)}}
\end{equation}
and,
\begin{equation}
\label{eq16}
  U_2(p_2)= \frac{1-e^{-iNkd\sin(\theta_2)}}{1-e^{-ikd\sin(\theta_2)}}
\end{equation}

\begin{figure}[h!] 
\vspace{0cm} 
\centerline{\includegraphics
[height=10cm,angle=-90.0]
{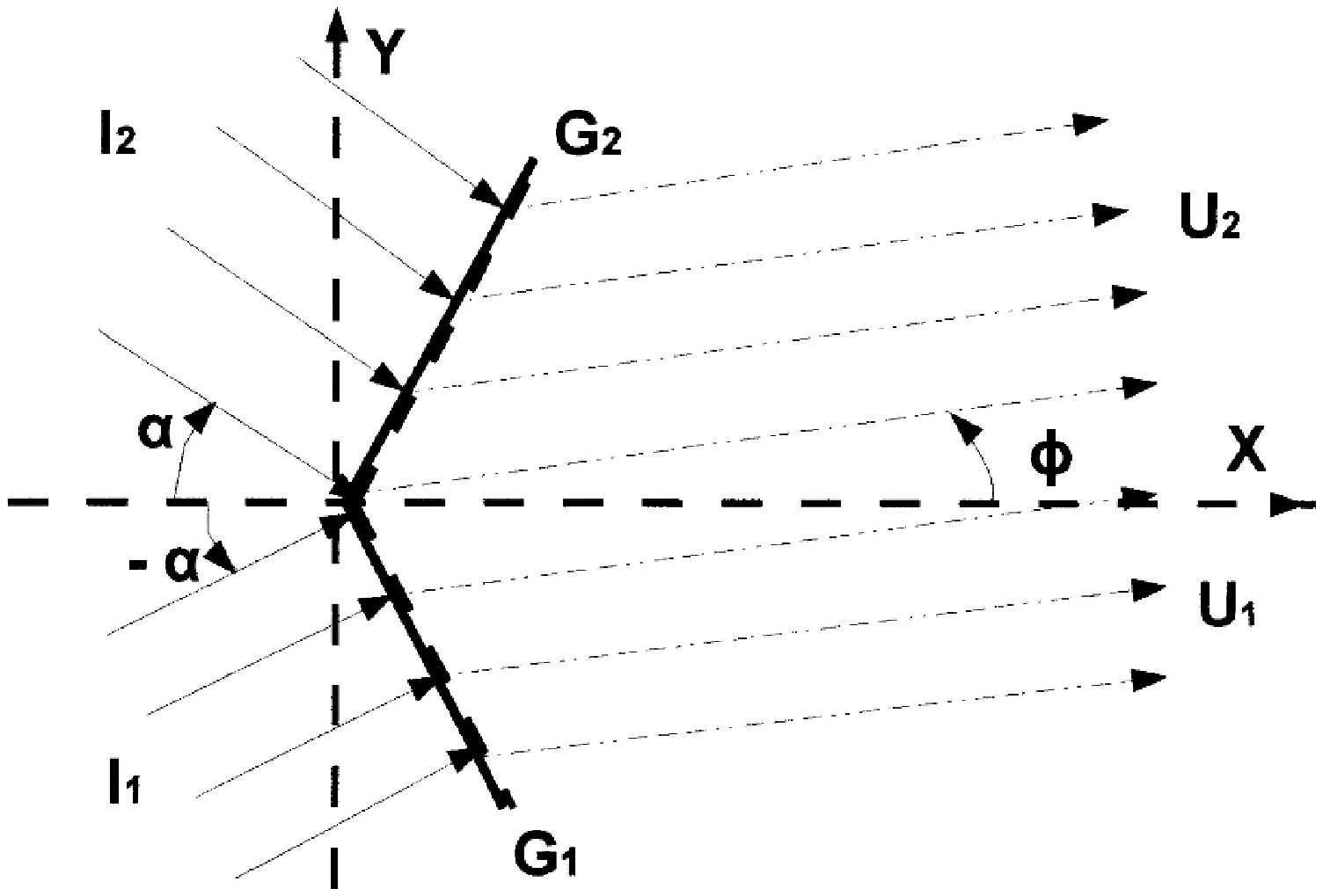}} 
\vspace{0cm} 
\caption{Two monochromatic coherent plane waves {\bf $I_1$} and {\bf $I_2$}, with their 
directions of motion on the $X-Y$ plane, and making angles $\alpha$, and $-\alpha$ 
with the $X$ axis, are each normally incident on identical diffraction gratings 
{\bf $G_1$} and {\bf $G_2$}, giving rise to diffracted amplitudes $U_1$, and 
$U_2$, to be observed at a large distance, in directions indicated by the angle 
$\phi$. } 
\end{figure}

In what follows we will only consider the case $\theta_1$ close to $-\theta_2$ 
(and close to $\alpha$). We have on this account, and for simplicity, set 
$U^{(0)}_1(p_1)=U^{(0)}_2(p_2)=1$, as they are, in general, slowly changing 
factors, depending only on the angles $\theta_1$ and $\theta_2$ respectively. 
We have also dropped the factor $e^{i\omega t}$, but have included a possible 
phase difference $e^{i\varphi}$, between the two beams.

The resulting amplitude a long distance from the gratings, can now
be computed as a function of the angle $\phi$ between the direction
of the diffracted beams and the $X$ axis (see Figure 2). We have,
\begin{eqnarray}
\label{eq17}
\theta_1 & = & \phi-\alpha \\
\theta_2 & = & \phi+\alpha \nonumber
\end{eqnarray}
and, therefore, the total amplitude $U$ as a function of $\phi$,
$\varphi$ and $k$ is given by,
\begin{equation}
\label{eq18}
  U(\phi,\varphi,k)= e^{i\varphi}
 \frac{1-e^{-iNkd\sin(\phi-\alpha)}}{1-e^{-ikd\sin(\phi-\alpha)}}
 +\frac{1-e^{-iNkd\sin(\phi+\alpha)}}{1-e^{-ikd\sin(\phi+\alpha)}}
\end{equation}
and the diffracted intensity is given by $I(\phi,\varphi,k)=
|U(\phi,\varphi,k)|^2$.

We are especially interested in the effect of changes in the wavelength on the 
pattern of diffracted light. Let us fix our attention on a particular 
wavelength $\lambda_0$. Define $k_0= 2\pi/\lambda_0$, and $k=2 \pi/\lambda$, 
for $\lambda \neq \lambda_0$, and assume that the angle $\alpha$ satisfies the 
relation,
\begin{equation}
\label{eq19}
  k_0 d \sin(\alpha)= \frac{2 \pi}{\lambda_0} d \sin(\alpha) =
  2\pi
\end{equation}
so that for $\lambda=\lambda_0$ the maximum intensity of both diffracted beams 
corresponds to $\phi=0$. For $\lambda \neq \lambda_0$ the condition 
(\ref{eq19}) implies that in the direction $\phi=0$ we have,
\begin{equation}
\label{eq20}
 I(0,\varphi,\frac{2\pi}{\lambda})=\left|  \frac{
1-{{\rm e}^{{\frac {-2i\pi {\lambda_0}N}{\lambda}}}} } { 1-{{\rm
e}^{{\frac {-2 i \pi {\lambda_0}}{\lambda}}}} } +{{\rm e}^{i\varphi}}
\frac{ 1-{ {\rm e}^{{\frac {2i\pi {\lambda_0}\,N}{\lambda}}}} }
 { 1-{{\rm e}^{{\frac {2i\pi {\lambda_0}}{\lambda}}}}
 } \right|^2
 \end{equation}

We consider now the two extreme cases $\varphi=0$, and
$\varphi=\pi$.

In the first case, ($\varphi=0$), for $\lambda$ close to
$\lambda_0$, we have,
\begin{equation}
\label{eq21}
 I(0,0,2\pi/\lambda)\simeq 4 N^2 -\frac{16\pi^2N^4} {3}
   \frac{(\lambda-\lambda_0)^2}{\lambda_0^2}
 +{\cal{O}}\left(\frac{(\lambda-\lambda_0)^3}{\lambda_0^3}\right)
 \end{equation}

This implies that the second term on the right of (\ref{eq21}) will
be of the same order as the first for $
(\lambda-\lambda_0)/\lambda_0 \sim 1/N$, and, therefore, we expect
that this will be also the order of the minimum observable
difference $\Delta \lambda/\lambda_0$. This is similar to the case
of two incoherent beams.

In the second case, ($\varphi=\pi$), for $\lambda$ close to
$\lambda_0$, we have,
\begin{equation}
\label{eq22}
 I(0,\pi,2\pi/\lambda)\simeq 4\pi^2N^4
  \frac{ (\lambda-\lambda_0)^2}{\lambda_0^2}
 -8\pi^2N^4
  \frac{(\lambda-\lambda_0)^3}{\lambda_0^3}
 +{\cal{O}}\left(\frac{(\lambda-\lambda_0)^4}{\lambda_0^4}\right)
 \end{equation}
But now we have that $I(0,\pi,2\pi/\lambda)=0$ for $\Delta \lambda=0$, and that 
for $\Delta \lambda/\lambda_0 \sim 1/N^2$ the first term in (\ref{eq22}) will 
be of order one, while the second is of order $1/N^2$. Therefore, in the case 
$\varphi=\pi$, the intensity at $\phi=0$ increases quadratically from zero to 
order one as $\Delta \lambda/\lambda_0$ increases from zero to order $1/N^2$. 
This would indicate that in this case we have a much larger sensitivity to 
changes in $\lambda$, and corresponding changes in $c$, than in the case of 
incoherent differences in wavelength. In the next Section we consider several 
examples to illustrate these points.

\section{Some examples.}

In the previous Section we considered only the intensities for $\phi=0$. In 
this Section we consider several explicit examples, where we assume that the source 
has a unique, highly stable frequency, assign definite values to the grating 
parameters, and compute the intensities as a function of $\phi$, for different 
choices of $\varphi$, and of $\Delta \lambda/\lambda_0$. Our choice of 
parameters is: $\alpha=\pi/6$, $N=10000$, $d=10000$ \text{\AA}, and we also 
take $\lambda_0=5000$\text{\AA}, in correspondence with (\ref{eq19}). The 
results are shown in several plots.

\subsection{The case $\varphi=0$}

\begin{figure}[h!] 
\centerline{\includegraphics
[height=6cm,angle=0.0]
{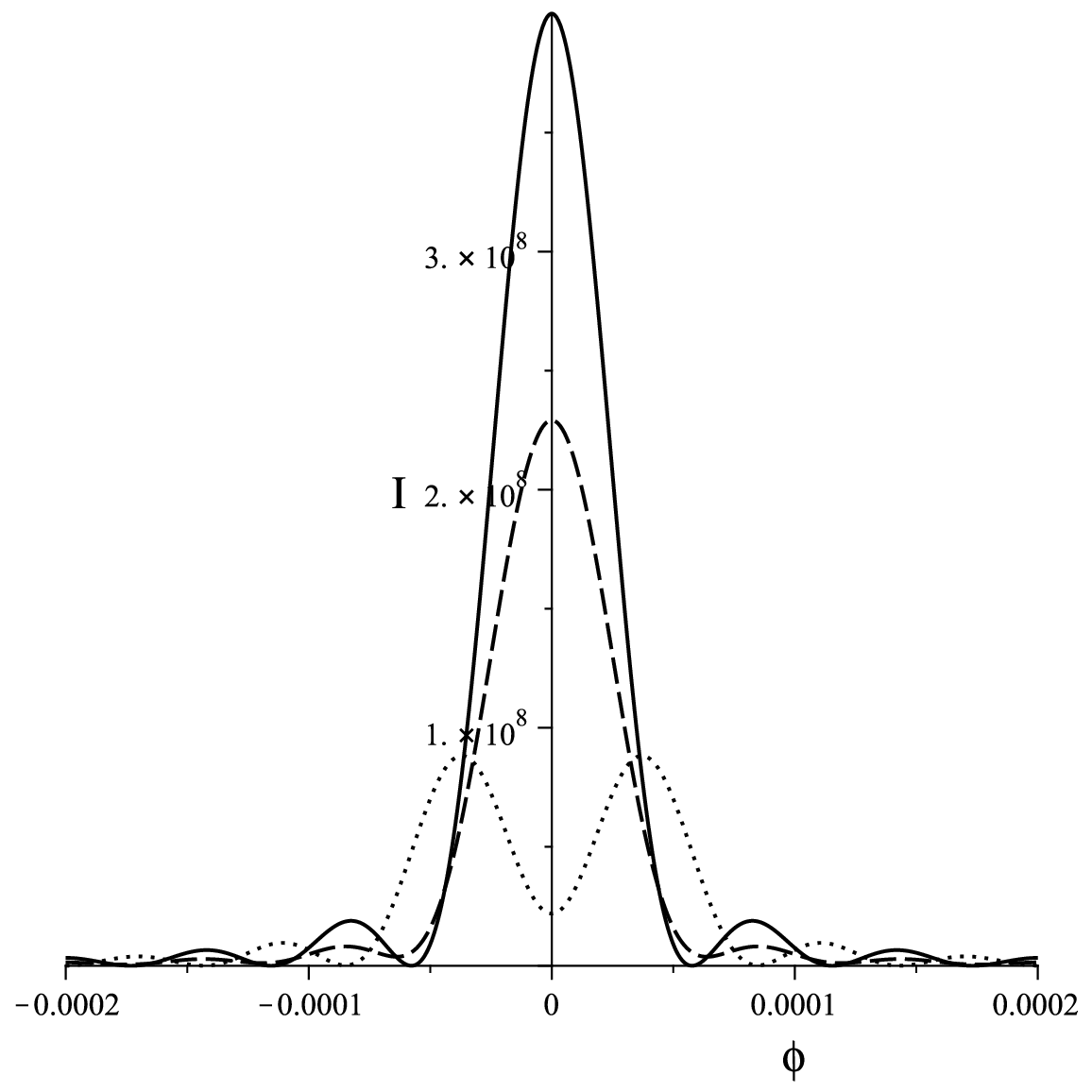}}
\vspace{1.5cm}
\caption{The case $\varphi=0$. Plots of the intensity $I$ as a function of 
$\phi$. The solid line corresponds to $\Delta \lambda=0$. The dashed line to 
$\Delta \lambda/\lambda_0=2\times 10^{-5}$. The dotted line to $\Delta 
\lambda/\lambda_0=4\times 10^{-5}$ }
\end{figure}

We consider first the case $\varphi=0$. In Figure 3, we have plots of the 
intensity as a function of $\phi$ for $\Delta \lambda/\lambda_0=0$, (solid 
curve), $\Delta \lambda/\lambda_0=2\times 10^{-5}$, (dashed curve) and  $\Delta 
\lambda/\lambda_0=4\times 10^{-5}$ (dotted curve). Although the dotted curve, 
($\Delta \lambda/\lambda_0=4\times 10^{-5}$), shows a clear separation of the 
maxima, we notice that already for the dashed curve, i.e. for $\Delta 
\lambda/\lambda_0=2\times 10^{-5}$, the maximum of the intensity has decreased 
to about half of that for $\Delta \lambda=0$. In fact, although not shown in 
Figure 3, we already have a decrease in the maximum of about $20 \%$, for 
$\Delta \lambda/\lambda_0=10^{-5}$, so that we may consider that in this case 
we have a resolution of the order of $10^{-5}$. We should also keep in mind 
that, for this choice of parameters, the maximum intensity is of the order of 
$10^8$, and that the width of the diffracted line, taken as the angular 
separation between the zeros next to the maximum, is about $\Delta \phi \simeq 
0.0001$.
 
In the next Subsection we consider the other extreme case, namely $\varphi=\pi$.

\subsection{The case $\varphi=\pi$}

In the case $\varphi=0$, and for $\Delta \lambda =0$,  we have complete 
cancellation of the diffracted amplitude only for $\phi=0$. For $\phi\neq 0$, 
there is, in general, a non vanishing intensity as shown in Figure 4. We 
notice, however, that the first maximum of the diffracted intensity at either 
the right or left of $\phi=0$, (with an intensity of the order of $700$, i.e., 
five orders of magnitude below that for $\Delta \lambda=0$, and $\varphi=0$), occur at 
$\phi \simeq \pm 0.005$, that is, at an angular separation about $100$ times 
the width of the diffracted line for $\varphi=0$ (notice that the plot extends 
from $\phi=-0.02$ to $\phi=0.02$). Thus, in the case $\varphi=\pi$, we have, 
around $\phi=0$, a region much larger than the width of the diffracted line, 
where the intensity for $\Delta \lambda=0$ can be taken as, essentially, zero. 
This is the situation depicted in Figure 5, which is restricted to the range 
$-0.0003 \leq \phi \leq +0.0003$.

\begin{figure}[h!] 
\centerline{\includegraphics
[height=7cm,angle=0.0]
{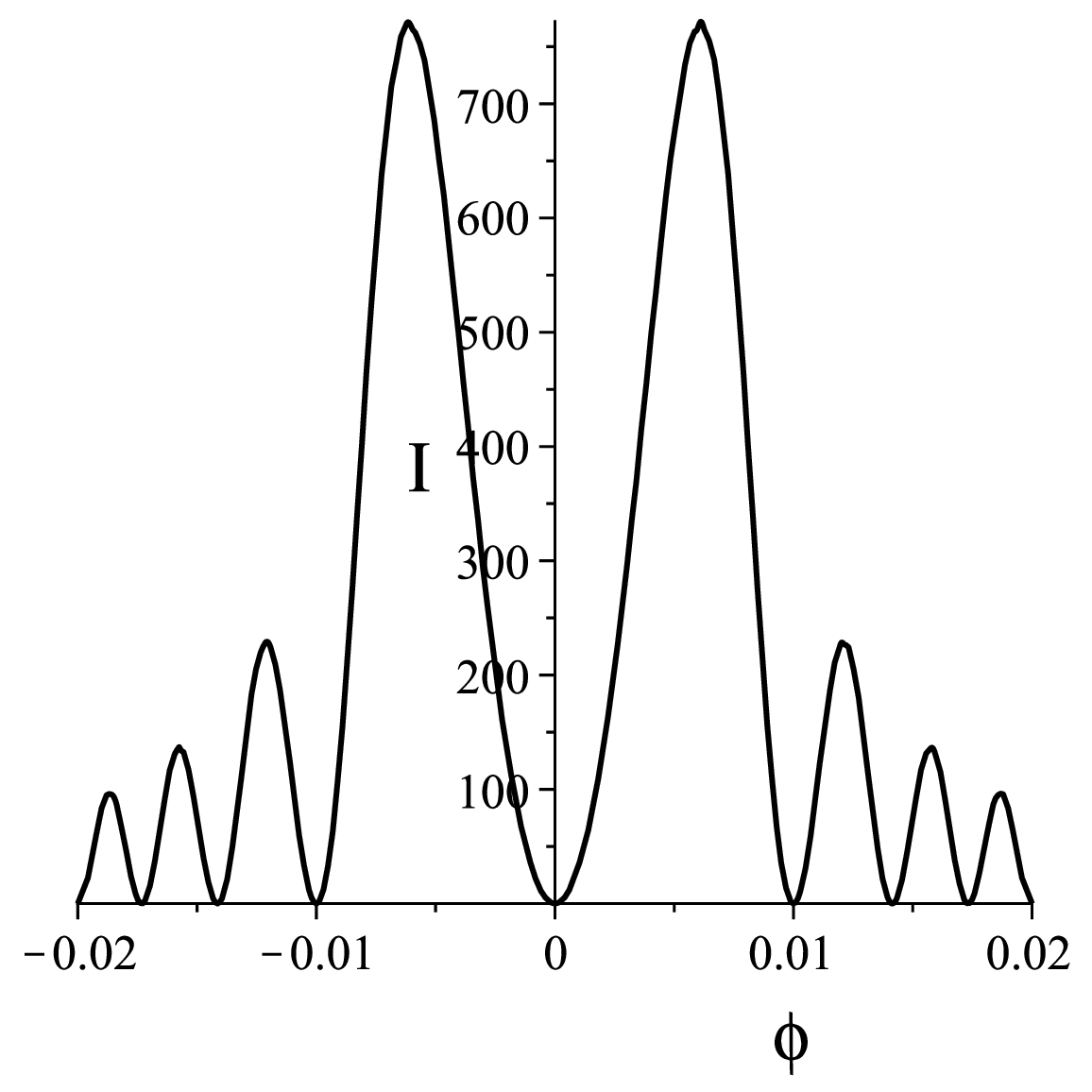}}
\vspace{1.5cm}
\caption{The case $\varphi=\pi$. Plot of the intensity $I$ as a function of 
$\phi$ for  $\Delta \lambda=0$. Notice that the range of $\phi$ is $-0.02 
\leq \phi \leq +0.02$, and that the maxima of $I$ are of the order of $700$.}
\end{figure}

In Figure 5 we have a plot of the intensity as a function of $\phi$ in the 
range $-0.0003 \leq \phi \leq +0.0003$ for several choices of $\Delta \lambda$. 
In accordance with the previous discussion, the curve corresponding to $\Delta 
\lambda=0$, plotted as a dotted line, can barely be distinguished from the 
horizontal axis, corresponding to $I=0$. The dashed line corresponds to $\Delta \lambda/
\lambda_0=4\times 10^{-8}$, and the solid line to $\Delta \lambda/\lambda_0=8\times 
10^{-8}$. We can see that in this case the arrangement displays a sensitivity 
to changes in $\lambda$, and, therefore in $c$, of the order of 1 part in 
$10^{-8}$. This, however, requires $\varphi=\pi$ rather accurately. In the next 
Subsection we consider a value of $\varphi$ close to but not equal to $\pi$.

\begin{figure}[h!] 
\centerline{\includegraphics
[height=8cm,angle=0.0]
{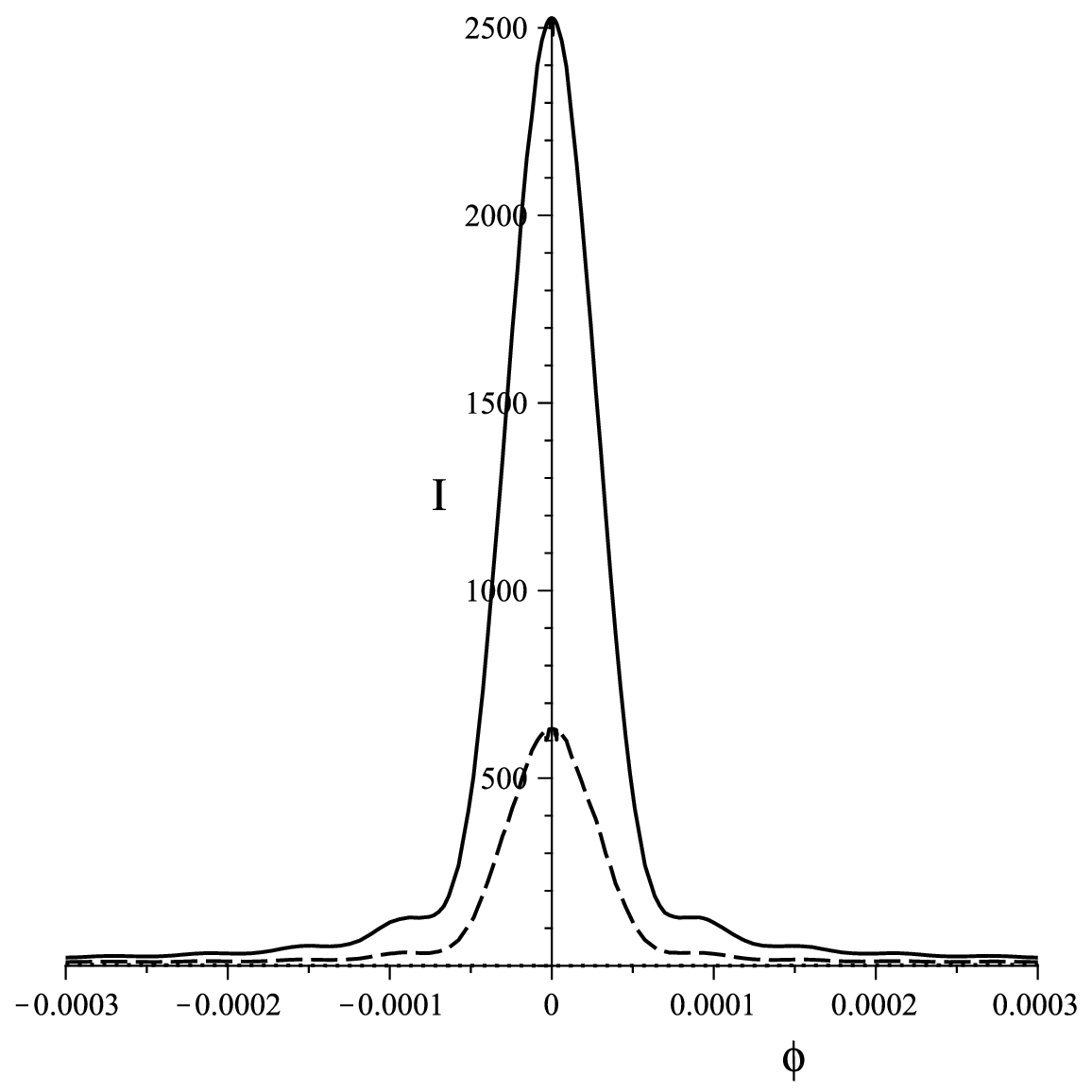}}
\vspace{1.5cm}
\caption{The case $\varphi=\pi$. Plots of the intensity $I$ as a function of 
$\phi$. The dotted line (which on this scale cannot be separated from the 
horizontal axis) corresponds to $\Delta \lambda=0$. The dashed line to $\Delta 
\lambda/\lambda_0=4\times 10^{-8}$. The solid line to $\Delta 
\lambda/\lambda_0=8\times 10^{-8}$. }
\end{figure}

\subsection{The cases $\varphi\neq\pi$, but with $\varphi$ close to $\pi$.}

Consider again (\ref{eq18}), with $k=k_0$, and $\alpha$ satisfying condition 
(\ref{eq19}). Expanding in $\phi$ up to order $\phi^4$, after some 
simplifications, and keeping in each term the leading order in $N$, we arrive 
at the following expression,
\begin{eqnarray}
\label{Idephi}
I(\phi,\varphi,k) & \simeq & 2A{N}^{2}+\frac{{N}^{4}}{6} \left( -BA+12\,\sin \left( \varphi \right) 
\pi  \right) {\phi}^{2} \\
 & + & {\frac {N^4}{360}} \left( 2 {B}^{2}A{N
}^{2}-2B \left( 30 \pi \sin \left( \varphi \right)  -BA \right) N+CA+
360{\pi }^{2} \right) {\phi}^{4}+{\cal{O}} \left( {\phi}^{6} \right)   \nonumber
\end{eqnarray}
where,
\begin{eqnarray}
\label{Idephi2}
A & = &  1  + \cos(\varphi) \nonumber \\
B & = &   k^2 d^2 -4 \pi^2 \\
C & = &  \left( 20+24{\pi }^{2} \right) B+48{\pi }^{4}-3{k}^{4}{d}^{4}-420{\pi }^{2} \nonumber
\end{eqnarray}

In the case $\varphi=0$ (\ref{Idephi}) reduces to,
\begin{equation}
\label{Idephi4}
I(\phi,\varphi,k)\simeq 4 N^2- \frac{B N^4}{3} \phi^2 +{\cal{O}}\left(\phi^4 \right) ,
\end{equation}
so that, as already indicated, for $\phi=0$ the intensity has a maximum of 
intensity, in this case, of the order of $N^2$, while for $\varphi=\pi$ the 
first two terms on the right of (\ref{Idephi}) (of order $N^2$ and $N^4$, 
respectively) have vanishing coefficients, and we are left with,
\begin{equation}
\label{Idephi6}
I(\phi,\varphi,k)\simeq \pi^2 N^4 \phi^4 +{\cal{O}}\left(\phi^6 \right).
\end{equation} 

We notice, however, that for $\varphi \neq \pi$, an effective cancellation of 
the first term on the right of (\ref{Idephi}) would require that 
$1+\cos(\varphi)$ be of the order of $1/N^2$ or less, which, in turn, requires 
$|\varphi-\pi|$ of the order of $1/N$ or less. On this account we consider now 
what happens for $\varphi$ close to $\pi$, but such that the first term in 
(\ref{Idephi}) is still dominant.

As a first example we chose the value  $\varphi=0.95 \pi$ which, on account of 
the previous discussion, is rather far from the value required for complete 
cancellation of the leading terms in (\ref{Idephi}). The resulting intensities are 
shown in Figure 6 for several choices of $\Delta \lambda/\lambda$. It is clear 
from this plot that already for $\Delta \lambda/\lambda_0=1\times 10^{-6}$ 
there is a substantial change in the intensity.

\begin{figure}[h!] 
\centerline{\includegraphics
[height=8cm,angle=0.0]
{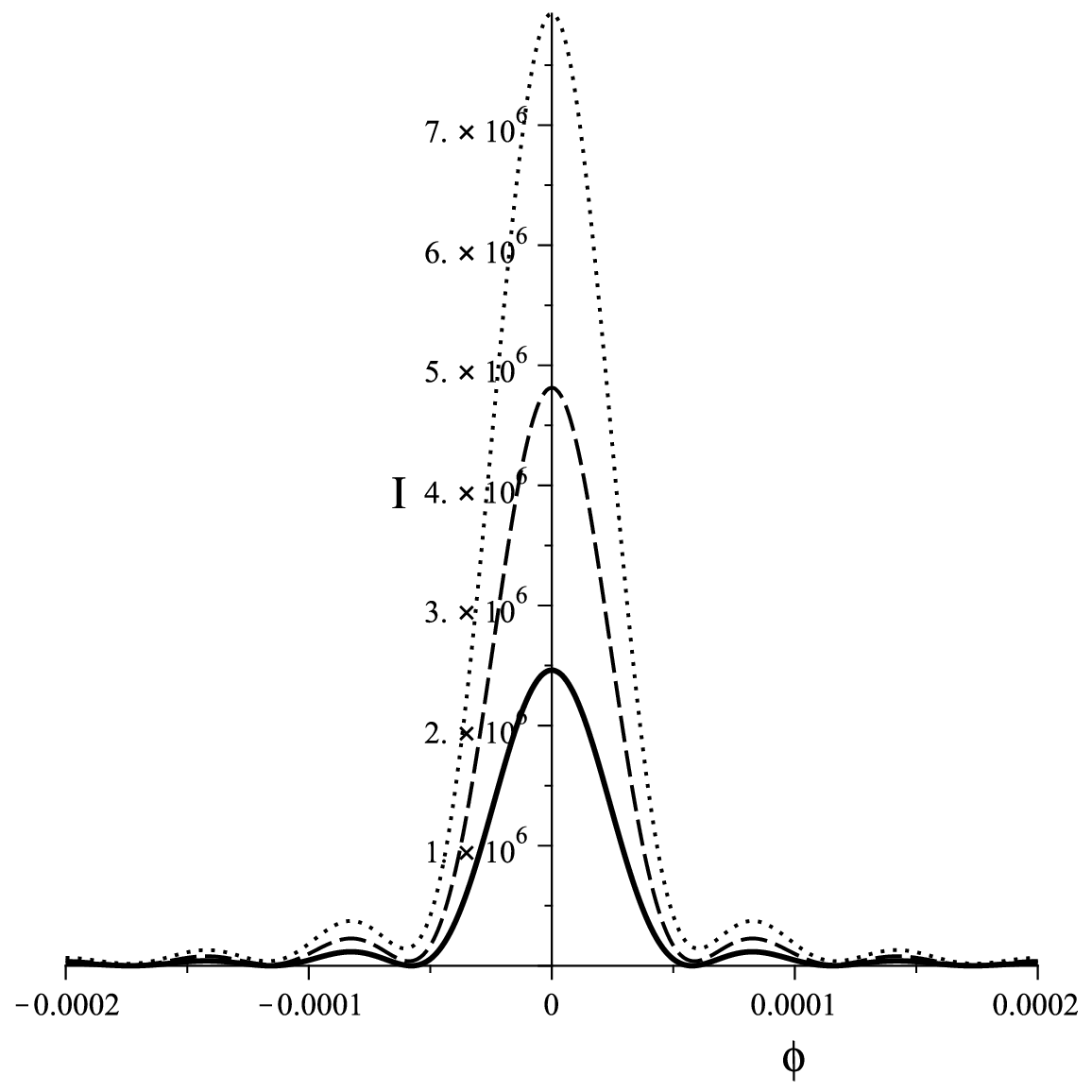}}
\vspace{1.5cm}
\caption{The case $\varphi=0.95 \pi$. Plots of the intensity $I$ as a function of 
$\phi$. The solid line corresponds to $\Delta \lambda=0$. The dashed line to $\Delta 
\lambda/\lambda_0=1\times 10^{-6}$. The dotted line to $\Delta 
\lambda/\lambda_0=2\times 10^{-6}$. }
\end{figure}

As a second example we considered  $\varphi=0.97 \pi$. The results are shown in 
Figure 7. This, as expected, shows a higher sensitivity to changes in 
$\lambda$. In fact, already for $\Delta \lambda/\lambda_0=2.8\times 10^{-7}$ we 
see a change of about $50 \%$ in the line intensity. 

\begin{figure}[h!] 
\centerline{\includegraphics
[height=8cm,angle=0.0]
{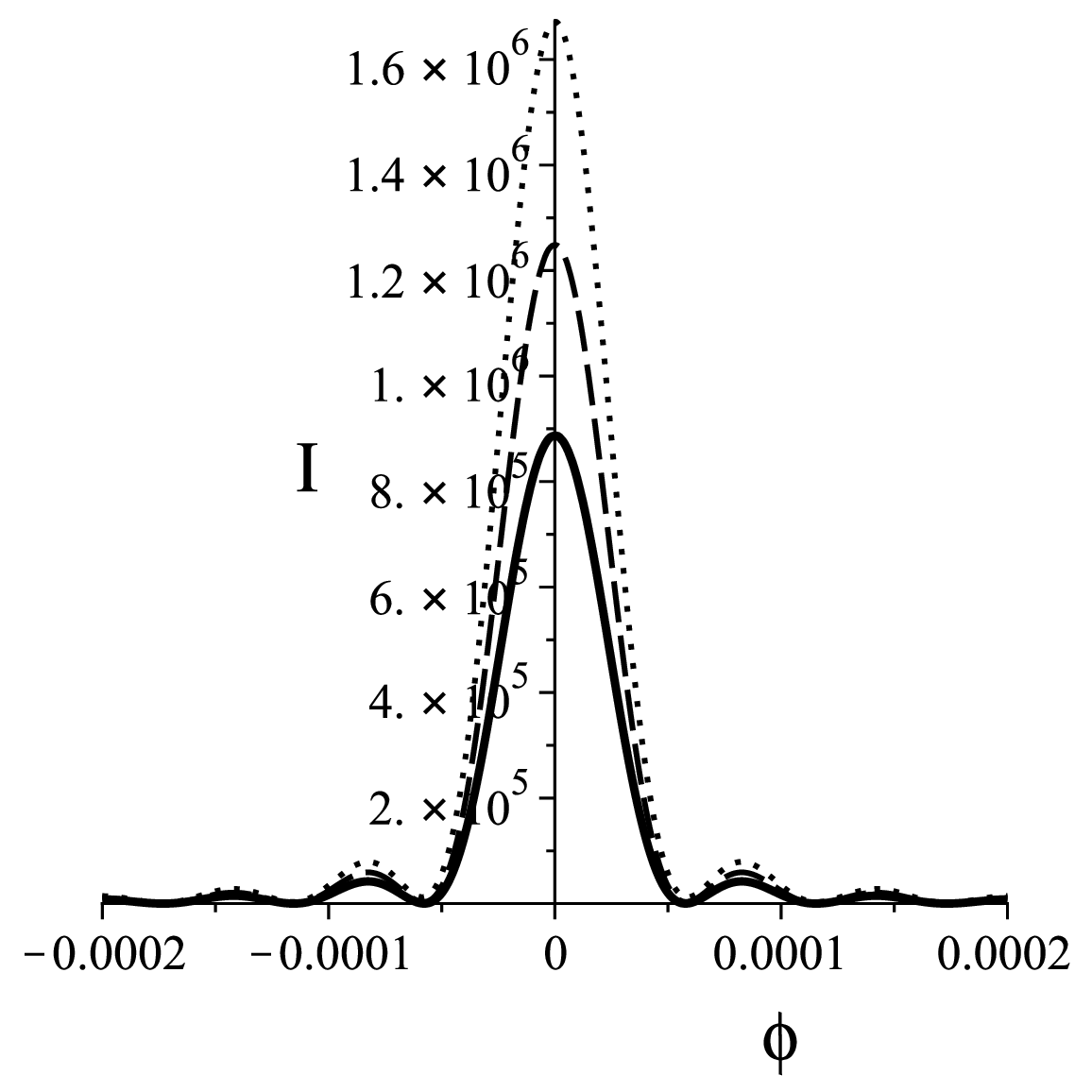}}
\vspace{1.5cm}
\caption{The case $\varphi=0.97 \pi$. Plots of the intensity $I$ as a function of 
$\phi$. The solid line corresponds to $\Delta \lambda=0$. The dashed line to $\Delta 
\lambda/\lambda_0=2.8\times 10^{-7}$. The dotted line to $\Delta 
\lambda/\lambda_0=5.6\times 10^{-7}$. }
\end{figure}

We close this Section mentioning, although we do not provide explicit examples, 
that an increase in $N$ results also in an increase in sensitivity of the 
setup. In fact definitive numbers should better refer to a particular 
experimental arrangement. In the next Section we sketch, but only as an 
illustrative example, a possible arrangement, without entering in any 
constructive detail.

\section{An experimental setup}

A possible basic setup for an experiment using the previous results can be 
described as follows (see Figure 8). We first assume an auxiliary orthogonal 
coordinate system with axes $(X,Y,Z)$, (not shown in the figure), with X and Y 
in the plane of the figure, and Z perpendicular to that plane, and place a 
frequency stabilized laser ($\bf{La}$ in the figure) at the origin of 
coordinates. The laser emits monochromatic light along the $X$ axis. An 
artifact, $\bf{B}$, splits the laser light into two equal but separate beams 
that are directed to slits $\bf{S_1}$, and $\bf{S_2}$. A pair of lenses 
$\bf{L_1}$, and $\bf{L_2}$, situated in such a way that the slits are in their 
focal plane, transform the beams emerging from the slits into two plane waves. 
These are reflected on mirrors $\bf{M_1}$, and $\bf{M_2}$, so that they 
eventually arrive at the gratings $\bf{G_1}$, and $\bf{G_2}$, that have their 
grooves parallel to the direction Z, exactly in the form indicated in Figure 2, 
so that the previous discussion and results apply to the emerging diffracted 
light. The corresponding Fraunhoffer pattern can then be observed on the screen 
${\bf S}$, in the focal plane the lens ${\bf L_3}$. The whole apparatus could be 
set on a platform, so that it can be oriented in different directions. Several 
comments are in order regarding this setup. The arrangement should include (not 
shown in Figure 8) some way of controlling both the amplitude and the phase of 
the beams in their way to the gratings $\bf{G_1}$, and $\bf{G_2}$. This can be 
achieved in different ways, and we have left it out for clarity. Regarding a 
concrete setup, besides the usual precautions, such as vibration isolation, a 
main concern would be temperature control, because a change in temperature, 
through temperature dilation, would result in a change in $d$, and this is 
equivalent to a change in $\lambda$. Similar regards concern mechanical 
stability, because changes in the dimensions could introduce changes in the 
paths of light previous to the diffraction gratings, and this, in turn, changes 
in $\varphi$ and the angle $\alpha$. It would also be appropriate to enclose 
the apparatus in a vacuum, and isolate it from electromagnetic disturbances. 

\begin{figure}[h!]
\vspace{-2cm}
\centerline{\includegraphics
[height=12cm,angle=-90.0]
{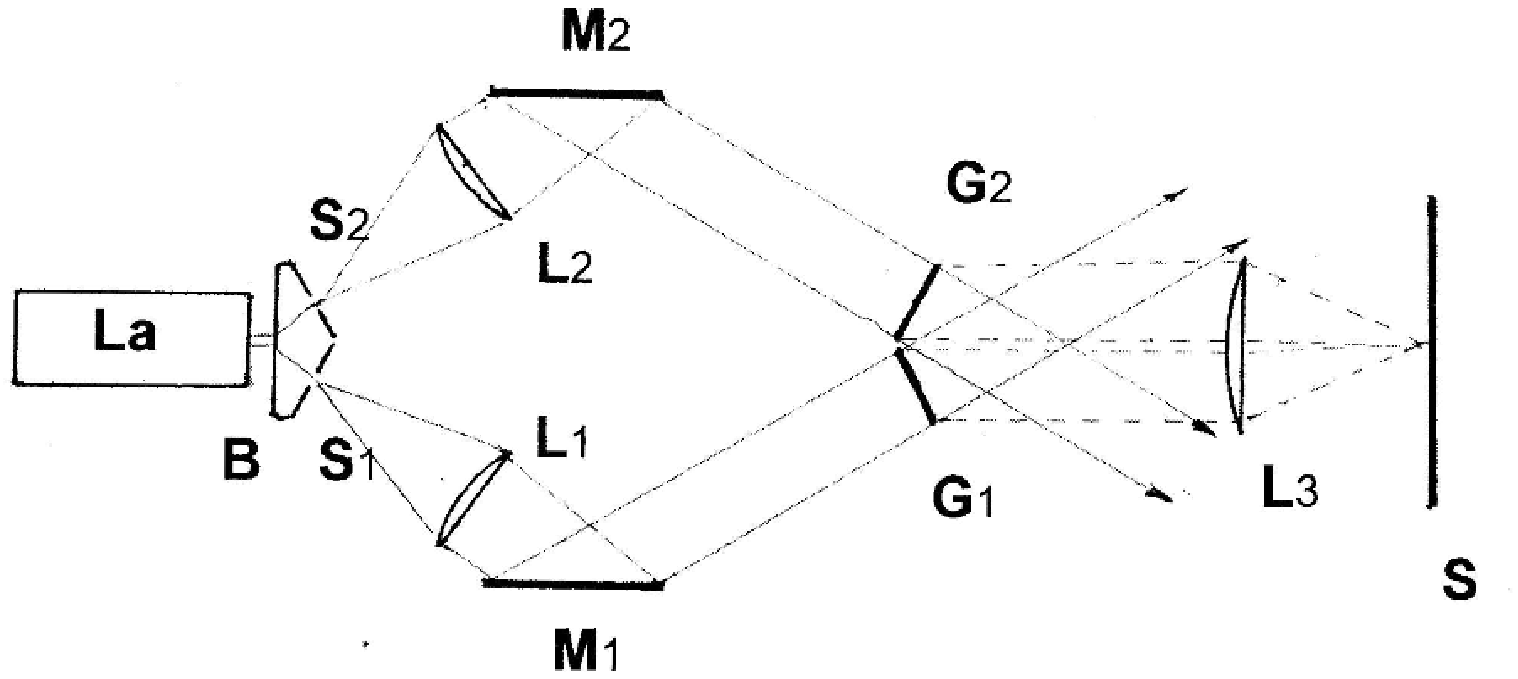}}
\vspace{-1.0cm}
\caption{The experimental setup.}
\end{figure}

In all this discussion we had in mind transmission diffraction gratings, but similar 
results would be obtained with reflective gratings, although the setup would 
be different.

\section{Comments}

In this note we have considered the problem of measuring the {\em one way} 
speed of light, taking into account the wave nature of light. For a 
monochromatic beam of light, propagating in a vacuum, this speed can be taken 
as the phase velocity, $c=\lambda/\tau$, where $\lambda$ is the wavelength and 
$\tau$ the period. As remarked already, these quantities can be measured 
independently. The period $\tau$ by a local clock which that does not involve 
any type of synchronization, and $\lambda$ by measuring the diffraction pattern 
generated as light passes through some known fixed structure, with no 
measurement of time or distance involved. We showed that although with an appropriate 
structure, such as a diffraction grating, one can measure $\lambda$ with a 
certain precision, we can use the properties of the diffraction gratings as 
regards wavelengths, to construct a setup that is sensitive to {\em changes} in 
wavelength, which, for fixed $\tau$, amounts to changes in $c$, to a much 
greater accuracy than that obtained by direct measurement of $\lambda$.  

The setup that we have considered can then be viewed as providing the basis 
for a {\em null} type experiment, where one checks for possible changes in the 
diffraction pattern as the apparatus is made to point in different directions. 
The invariance of the diffraction pattern (null result) would then provide an
upper bound on the possible anisotropy of the {\em one way} propagation of 
light. As shown here, this upper bound could be on the order of 1 part in 
$10^8$. Combining these results with the known bounds on the isotropy of the 
{\em two way} speed of light, which are much smaller than this, we would 
immediately obtain a value of the {\em one way} speed of light, with the same 
accuracy of the order of one part in $10^8$.

On the other hand, a definite non null result would establish the presence of 
an anisotropy in the one way speed of light, with an accuracy in the order of a 
few meters per second, whose origin and properties would certainly require much 
further analysis, both from the theoretical as from the experimental side.

\section*{Acknowledgments}

I am grateful to Eli Yudowsky for bringing the question of the {\em one way} speed of 
light to my attention.

\appendix

\section{A (slightly) more general setup.} 

In all the discussion in the text we have assumed equal amplitudes for the 
diffracted beams, as well as equal number of grooves ($N$) and equal groove 
separation ($d$). This led rather naturally in the case $\varphi=\pi$ to the 
total cancellation of the amplitude for $\phi=0$, as well as to a small value 
of the amplitude in an appropriate neighborhood of $\phi=0$.

In this Appendix we consider a slightly more general case, where we allow for 
different values of amplitudes, $N$ and $d$ for the gratings. We therefore 
write the total amplitude in the form,

\begin{equation}
\label{ap01}
U={\frac {1-{{\rm e}^{ik{ d_1}\,\sin \left( \phi-{ \alpha_1} \right) 
{ N_1}}}}{1-{{\rm e}^{ik{ d_1}\,\sin \left( \phi-{ \alpha_1}
 \right) }}}}+{\frac {{ a_2}\,{{\rm e}^{i\varphi}} \left( 1-{{\rm e}^{
ik{ d_2}\,\sin \left( \phi+{ \alpha_2} \right) { N_2}}} \right) }
{1-{{\rm e}^{ik{ d_2}\,\sin \left( \phi+{ \alpha_2} \right) }}}}
\end{equation}
with the intensity given by,
\begin{equation}
\label{ap02}
 I(\phi,\varphi,k)=|U|^2
\end{equation}

We fix now the angles $\alpha_i$ imposing the condition that for 
$\lambda=\lambda_0$, the maxima of the corresponding diffracted beams occur for 
$\phi=0$:
\begin{eqnarray}
\label{ap03}
\sin(\alpha_1) & = & \frac{1}{2 \pi k_0 d_1}=\frac{\lambda_0}{d_1} \\
\sin(\alpha_2) & = & \frac{1}{2 \pi k_0 d_2}=\frac{\lambda_0}{d_2}  \nonumber
\end{eqnarray}

Replacing (\ref{ap01}), and (\ref{ap03}) in (\ref{ap02}), after some 
simplifications we get,

\begin{equation}
\label{ap04}
I=\left|\frac{ 1-{{\rm e}^{{\frac {2\,i\pi \,{ N_1}\, \left( \sin \left( \phi 
\right) { dl_1}-\cos \left( \phi \right) { \lambda_0} \right) }{\lambda}}}} }  
{ 1-{{\rm e}^{{\frac {2\,i\pi \, \left( \sin \left( \phi \right) { dl_1}-\cos 
\left( \phi \right) { \lambda_0} \right) }{\lambda}}}} } +\frac{{ a_2}\,{{\rm 
e}^{i \varphi}} \left( 1-{{\rm e}^{{\frac {2\,i\pi \,{ N_2}\, \left( \sin 
\left( \phi \right) { dl_2}+\cos \left( \phi \right) { \lambda_0} \right) 
}{\lambda}}}} \right)}  { 1-{{\rm e}^{{\frac {2\,i\pi \, \left( \sin \left( 
\phi \right) { dl_2}+\cos \left( \phi \right) { \lambda_0} \right) 
}{\lambda}}}} }\right|^2
\end{equation}
where,
\begin{eqnarray}
\label{ap05}
dl_1 & = & \sqrt{d_1^2-\lambda_0^2} \\
dl_2 & = & \sqrt{d_2^2-\lambda_0^2}  \nonumber
\end{eqnarray}

We are particularly interested in the behaviour of $I$ for $\varphi=\pi$ and 
$\lambda=\lambda_0$, near $\phi=0$. To second order in $\phi$ we find,
\begin{eqnarray}
\label{ap06}
I(\phi,\varphi,2\pi/\lambda_0) & \simeq & \left( 2 a_2 N_1N_2 
(\cos(\varphi)+1)+(N_1-a_2N_2)^2\right) \\
    & & + \frac{ 2 \pi a_2 N_1 N_2 \left[ (N_1-1)dl_1-(N_2-1)dl_2\right] 
    \sin(\varphi)}{\lambda_0} \phi + {\cal{O}}(\phi^2) \nonumber
\end{eqnarray}

Therefore, the intensity will vanish for $\varphi=\pi$ and $\phi=0$ only if we 
impose:
\begin{equation}
\label{ap07}
 a_2=\frac{N_1}{N_2}
\end{equation}

If, assuming now (\ref{ap07}) and $\varphi=\pi$, we expand the intensity to 
order $\phi^3$, we find,
\begin{eqnarray}
\label{ap08}
I(\phi,\pi,2\pi/\lambda_0) & \simeq & \frac{\pi^2 
N_1^2\left[(N_1-1)dl_1-(N_2-1)dl_2\right]}{\lambda_0^2}\phi^2  \nonumber \\
    & & + \frac{\pi^2 
    N_1^2(N_1+N_2-2)\left[(N_1-1)dl_1-(N_2-1)dl_2\right]}{\lambda_0^2}\phi^3  
    \\
    & &  + {\cal{O}}(\phi^4) \nonumber
\end{eqnarray}
 
The term of order $\phi$ vanishes already for $\varphi=\pi$. The vanishing of 
the terms in $\phi^2$ and $\phi^3$ for $\varphi\neq\pi$ would require,
\begin{equation}
\label{ap09}
\frac{N_1-1}{N_2-1}=\frac{dl_2}{dl_1}= 
\frac{\sqrt{d_2^2-\lambda_0^2}}{\sqrt{d_1^2-\lambda_0^2}}
\end{equation}
but this cannot be satisfied in general for integer values of $N_1$ and $N_2$. 
Nevertheless, we may impose that, for integer values of $N_1$ and $N_2$, the 
left hand side of (\ref{ap09}) be a good numerical approximation to the right 
hand side of (\ref{ap09}). We shall assume that $N_1$ and $N_2$ have been 
chosen that way. 

Imposing now (\ref{ap07}), and assuming that (\ref{ap09}) is (approximately) satisfied, we find,
 
\begin{equation}
\label{ap10}
I \simeq {\frac {{\pi }^{2}{{  N_1}}^{2} \left( 4{\pi }^{2}{{  dl_2
}}^{2}{{  dl_1}}^{2} \left( dl_1-{  dl_2} \right) ^{2}+9\,{
{  \lambda_0}}^{4} \left( {  dl_1}+{  dl_2} \right) ^{2} \right) 
 \left({  N_1}-1 \right) ^{2}}{{{  \lambda_0}}^{4}{36{  dl_2}}^{2}}
}{\phi}^{4}+O \left( {\phi}^{5} \right) 
\end{equation}
which, for $d_1\simeq d_2$, and $N_1\simeq N_2 >>1$ reduces to,
\begin{equation}
\label{ap11}
I \simeq 4 \pi^2 N_1^4
{\phi}^{4}+O \left( {\phi}^{5} \right) 
\end{equation}
which is the same as (\ref{Idephi6}).

It is also interesting to see the general behaviour of the intensity for 
$\phi=0$, and $\lambda \neq\ \lambda_0$. Assuming again (\ref{ap07}) and 
$N_1\simeq N_2 >>1$ , we get, 
\begin{equation}
\label{ap12}
I \simeq {\frac {{\pi }^{2}{{  N_1}}^{2} \left( {  N_1}+{  N_2} \right) 
^{2}}{{{  \lambda_0}}^{2}}} \left( \lambda-{  \lambda_0} \right) ^{2}-
2\,{\frac {{\pi }^{2}{{  N_1}}^{2} \left( {  N_1}+{  N_2}
 \right) ^{2}}{{{  \lambda_0}}^{3}}} \left( \lambda-{  \lambda_0}
 \right) ^{3}+O \left(  \left( \lambda-{  \lambda_0} \right) ^{4}
 \right) 
\end{equation}
which reduces to (\ref{eq22}) for $N_1\simeq N_2 \simeq N$.

The derivations in this Section indicate that with appropriate fixing of the 
relative amplitudes ($a_2$) and number of grooves illuminated ($N_1,N_2$), we 
essentially recover the results obtained in the main text, regarding the 
properties of the total diffracted intensity in the case $\varphi=\pi$. In the 
next Section we consider an explicit numerical example to illustrate this 
points.

\section{A numerical example}

In this Section we shall consider as an explicit example, for $\varphi=\pi$, 
the case where  $d_1=10000$ \text{\AA}, and  $d_2=10010$ \text{\AA}. We also 
take $\lambda_0=50000$ \text{\AA}. 

Next we consider (\ref{ap09}), which in this reads,
 
\begin{equation}
\label{ap13}
\frac{N_1-1}{N_2-1}= 1.001333...
\end{equation}

Then, taking, for instance, $N_1=10000$, if choose $N_2=9987$ we have,
\begin{equation}
\label{ap14}
\frac{N_1-1}{N_2-1}= 1.001301...
\end{equation}

Taking this into account, we choose $N_1=10000$, and $N_2=9987$, set $a_2= 
N_1/N_2$ and replace all numerical values in (\ref{ap04}) and (\ref{ap05}). 
Using the resulting expression we compute $I(\phi,\varphi,2 \pi/\lambda)$ for 
several choices of $\lambda$. In Figure 9 we have plot of $I$ as a function of 
$\phi$ for $\varphi =\pi$, and $\Delta \lambda=0$, in the range $-0.02 \leq 
\phi \leq 0.02$, which shows that this case of $d_1 \neq d_2$ can be made 
essentially equal to the case of $d_1 = d_2$ by appropriate choices of $a_2$ 
and $N_1,\; N_2$. In an experimental set up this means controlling the 
amplitudes of the beams previous to the gratings, and controlling the extent of 
the gratings illuminated by those beams.

\begin{figure}[h!] 
\centerline{\includegraphics
[height=7cm,angle=0.0]
{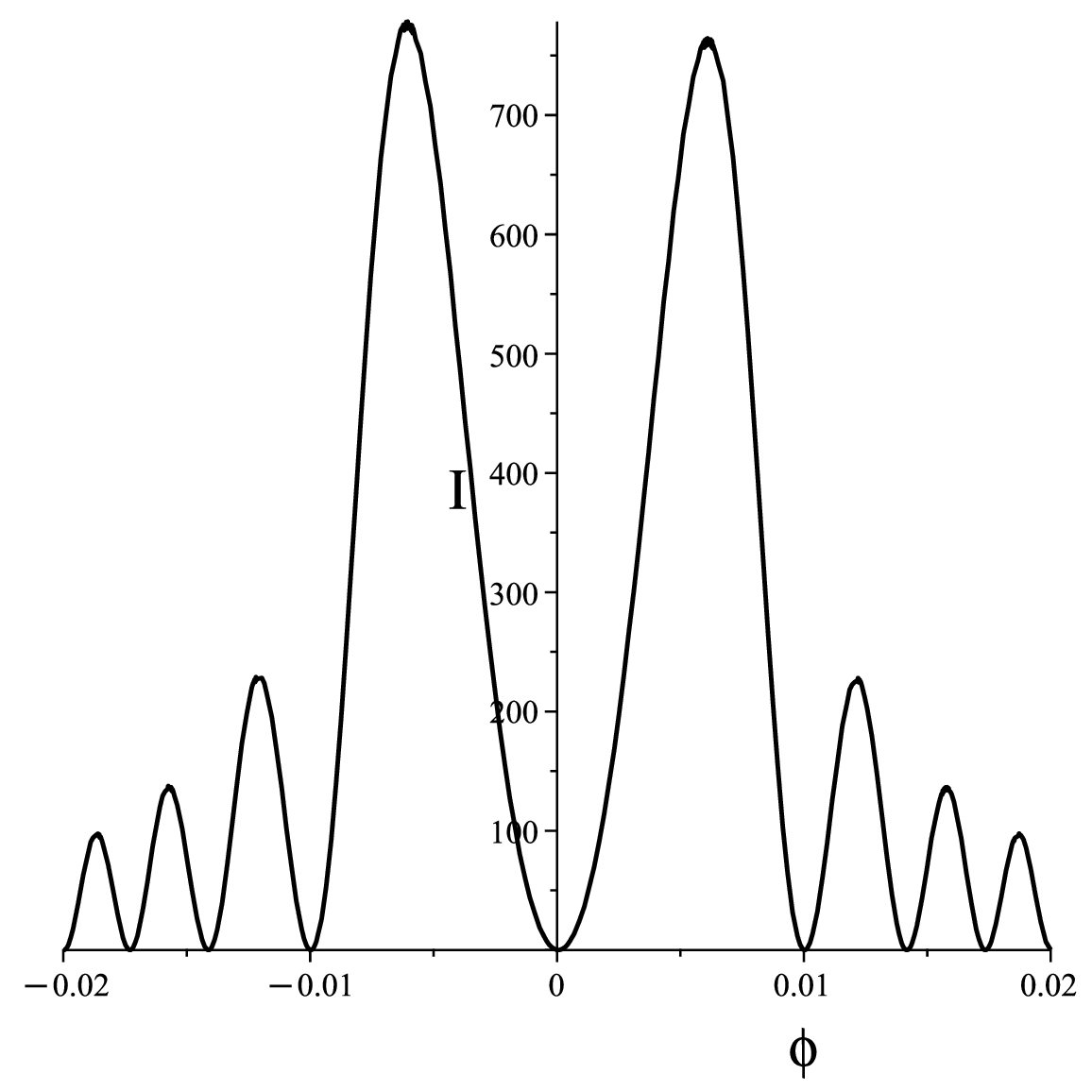}}
\vspace{1.0cm}
\caption{A case with $\Delta\lambda=0$ but with $d_1 \neq d_2$, described in the text, with 
$\varphi=\pi$, and appropriate choices of $a_2$ and $N_1,\; N_2$.}
\end{figure}

In Figure 10 we have a plot of the intensity for several values of $\lambda$, 
in the range $-0.0002 \leq \phi \leq 0.0002$. The solid curve, which shows a 
clear departure from the case $\lambda=\lambda_0$,  corresponds to 
$\lambda=5000.0003$, that is $\Delta \lambda/\lambda_0=6 \times 10^{-8}$, 
although this departure is apparent already for $\Delta \lambda/\lambda_0=2 
\times 10^{-8}$ (the dashed curve). The case $\Delta \lambda/\lambda_0=0$ is 
also shown in the figure as the dotted curve, barely distinguishable from the 
axis $I=0$. 

\begin{figure}[h!] 
\centerline{\includegraphics
[height=7cm,angle=0.0]
{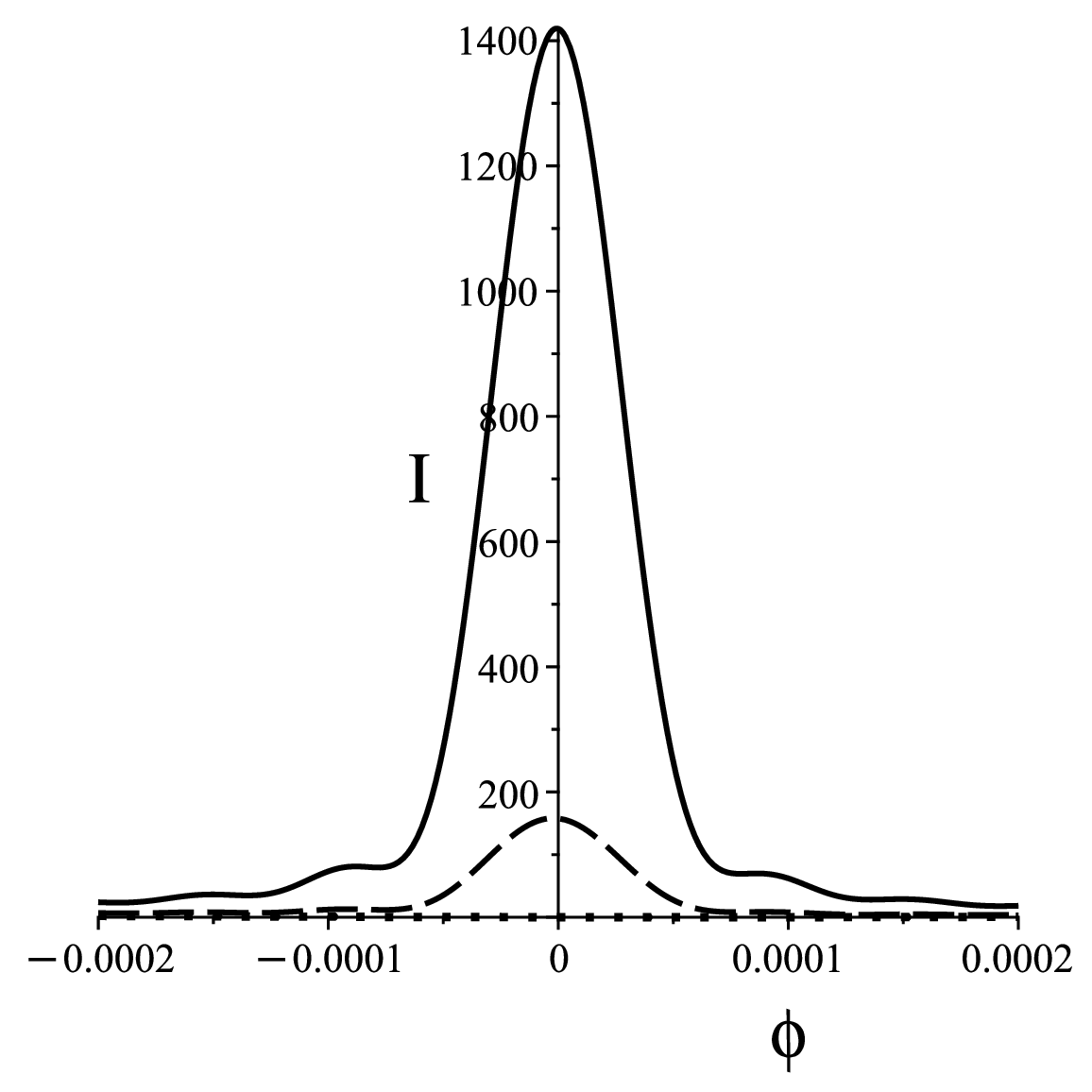}}
\vspace{1.0cm}
\caption{ Plots of the intensity $I$ as a function of $\phi$, in the range 
$-0.0002 \leq \phi \leq 0.0002$. The solid curve corresponds to $\Delta 
\lambda/\lambda_0=6 \times 10^{-8}$. The dashed curve to $\Delta 
\lambda/\lambda_0=2 \times 10^{-8}$. The dotted curve, corresponding to $\Delta 
\lambda/\lambda_0=0$ is barely distinguishable from the axis $I=0$. }
\end{figure}

In Figure 11, mainly for clarity, we show a plot of the case  $\Delta 
\lambda/\lambda_0=6 \times 10^{-8}$ of Figure 10, but here in the range $-0.002 
\leq \phi \leq 0.002$, that is ten times that of Figure 10, to show the clear 
separation of this ``signal'' from the ``background'' $\Delta 
\lambda/\lambda_0=0$.

\begin{figure}[h!] 
\centerline{\includegraphics
[height=7cm,angle=0.0]
{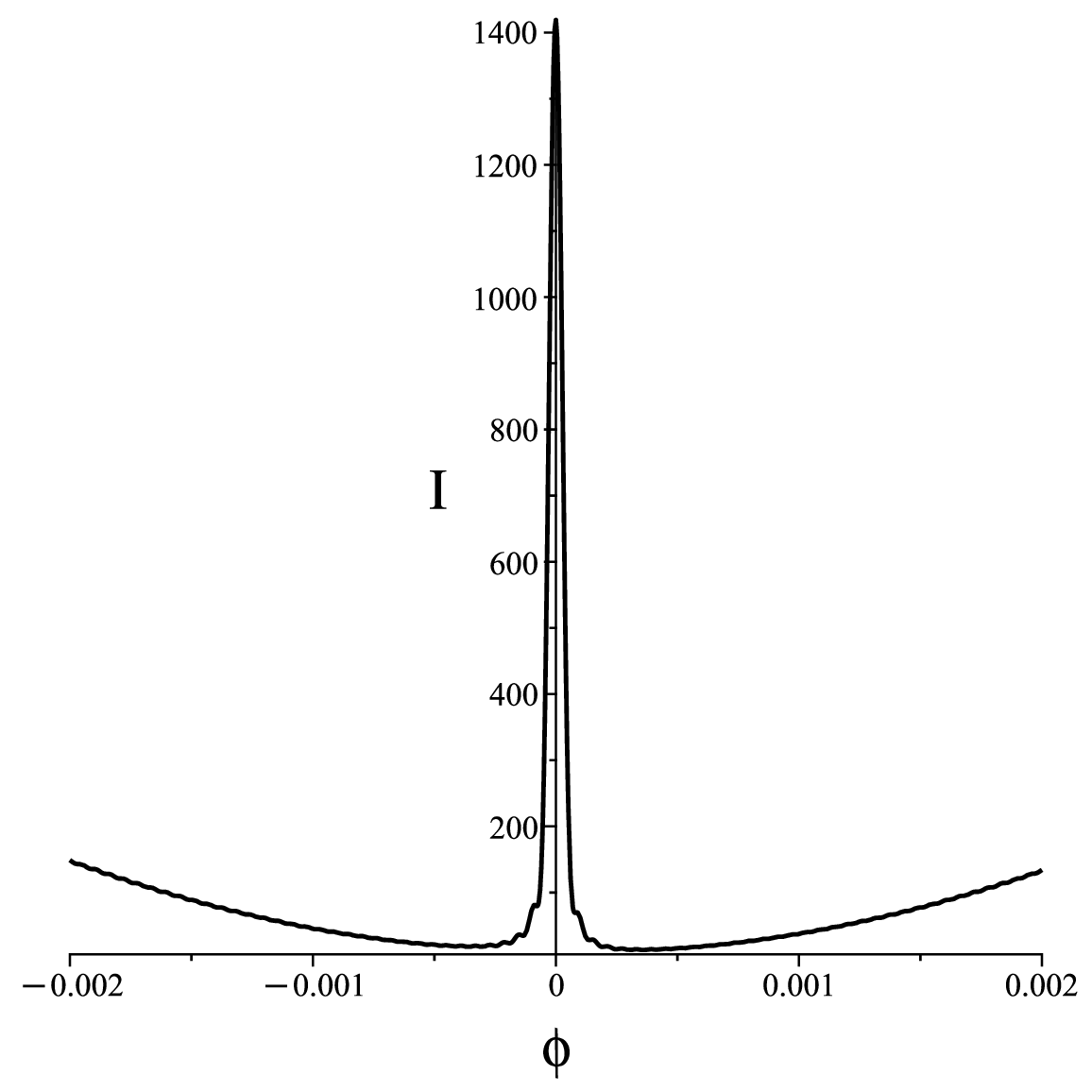}}
\vspace{1.0cm}
\caption{Plots of the intensity $I$ as a function of $\phi$, in the range 
$-0.002 \leq \phi \leq 0.002$, for $\Delta \lambda/\lambda_0=6 \times 
10^{-8}$.}
\end{figure}


\begin{thebibliography}{99}
\bibitem{refes} See for instance the review \cite{Anderson}. It is not our 
purpose here to add to this debate, rather, we will concentrate on the topic 
suggested by the title of this note.
\end{thebibliography}
\end{document}